\documentclass[10pt,format=sigconf]{acmart-cidr}
\AtBeginDocument{%
  \providecommand\BibTeX{{%
    \normalfont B\kern-0.5em{\scshape i\kern-0.25em b}\kern-0.8em\TeX}}}

\setcopyright{cidr}

\usepackage{tikz}
\usepackage{amsmath}
\usepackage{xspace}
\usepackage{graphicx}
\usepackage{listings}
\usepackage[most]{tcolorbox}
\usepackage[labelfont=small,textfont=small]{subcaption}
\usepackage{booktabs}
\usepackage{multirow}
\usepackage{tabularx}
\usepackage{paralist}
\usepackage{xcolor}
\usepackage{changepage}

\DeclareRobustCommand*\circledgray[1]{\tikz[baseline=(char.base)]{
            \node[shape=circle,draw=pink,inner sep=1pt,fill=lightgray,text=white,draw=gray, scale=0.8] (char) {#1};}}
            
\newcommand\sbullet[1][.5]{\mathbin{\vcenter{\hbox{\scalebox{#1}{$\bullet$}}}}}

\newif\ifcomments
\commentstrue
\ifcomments
    \providecommand{\jmh}[1]{{\protect\color{purple}{\bf [Joe: #1]}}}
    \providecommand{\alvin}[1]{{\protect\color{blue}{\bf [Alvin: #1]}}}
    \providecommand{\nc}[1]{{\protect\color{pink}{\bf [Natacha: #1]}}}
    \providecommand{\mpm}[1]{{\protect\color{violet}{\bf [Matthew: #1]}}}
    
\else
    \providecommand{\jmh}[1]{}
    \providecommand{\alvin}[1]{}
    \providecommand{\nc}[1]{}
    \providecommand{\mpm}[1]{}
\fi

\newcommand{\smallitem}[1]{\vspace{0.3em}\noindent\textbf{#1}}
\newcommand{\smallitembot}{\vspace{0.5em}\noindent}

\newcommand{\smallitemem}[1]{\vspace{0.3em}\noindent\emph{--- #1}}
\newcommand{\smallitemembot}{\vspace{0.5em}\noindent}

\newcommand{\sys}{\textsc{Hydro}\xspace}
\newcommand{\hydro}{\sys}
\newcommand{\logic}{\textsc{HydroLogic}\xspace}

\newcommand{\compiler}{\textsc{Hydrolysis}\xspace}
\newcommand{\runtime}{\textsc{Hydroflow}\xspace}
\newcommand{\lifter}{\textsc{Hydraulic}\xspace}

\lstdefinestyle{PythonStyle} {
basicstyle=\ttfamily\scriptsize,       
language=Python,
keywordstyle=\color{blue}\ttfamily\bfseries,
keywordstyle=[1]{\color{blue}\ttfamily\bfseries},
keywordstyle=[2]{\color{blue}\ttfamily\bfseries},
numbers=left,                   
numberstyle=\ttfamily\color{gray},      
stepnumber=1,                   
numbersep=8pt,                  
breakindent=0pt,
firstnumber=1,
showspaces=false,               
showstringspaces=false,         
showtabs=false,                 
frame=leftline,
tabsize=2,  		
captionpos=b,   		
breaklines=false,    	
breakatwhitespace=true,    
columns=fixed,
basewidth=0.52em,
numberblanklines=false,
escapechar=|,
morekeywords={}
}
\newtcblisting{pythonlisting}[2][]{
  enhanced,
  attach boxed title to top right={yshift=-3mm,xshift=-2mm},
  top = 0.1pt, bottom=0.1pt,
  colframe=black,
  colback=white,
  fonttitle=\bfseries\scriptsize, 
  colbacktitle=white,
  coltitle=black,
  boxed title style={
    boxrule=0pt,
    colframe=white,
    },
  listing only,
  listing style=PythonStyle,
  title=#2,
  #1}

\lstdefinestyle{HydroStyle} {
basicstyle=\ttfamily\scriptsize,       
language=Python,
keywordstyle=\color{blue}\ttfamily\bfseries,
numbers=left,                   
numberstyle=\ttfamily\color{gray},      
stepnumber=1,                   
numbersep=8pt,                  
breakindent=0pt,
firstnumber=1,
showspaces=false,               
showstringspaces=false,         
showtabs=false,                 
frame=leftline,
tabsize=2,  		
captionpos=b,   		
breaklines=false,    	
breakatwhitespace=true,    
columns=fixed,
basewidth=0.52em,
numberblanklines=false,
escapechar=|,
otherkeywords={:=,delete,update,sum},
keywordstyle=[1]{\color{blue}\ttfamily\bfseries},
keywordstyle=[2]{\color{blue}\ttfamily\bfseries},
keywordstyle=[3]{\color{red}\ttfamily\bfseries},
morekeywords=[1]{var,table,key,partition,handler,on,query,isolation,consistency,availability,failures,deployment,processor,default,latency,cost,merge,send,once,domain,target},
morekeywords=[3]{:=,delete,update,sum},
alsoother={@}
}
\newtcblisting[auto counter]{oldhydrolisting}[2][]{round corners, 
    top = 0.1pt, bottom=0.1pt,
    fonttitle=\bfseries\scriptsize, colframe=gray, listing only, 
    listing style=HydroStyle, 
    notitle,
    #1}
    
\tcbuselibrary{skins}

\lstdefinestyle{SimpleStyle} {
basicstyle=\ttfamily\scriptsize,       
language=Python,
numbers=left,                   
numberstyle=\ttfamily\color{gray},      
stepnumber=1,                   
numbersep=8pt,                  
breakindent=0pt,
firstnumber=1,
showspaces=false,               
showstringspaces=false,         
showtabs=false,                 
frame=leftline,
tabsize=2,  		
captionpos=b,   		
breaklines=false,    	
breakatwhitespace=true,    
columns=fixed,
basewidth=0.52em,
numberblanklines=false,
escapechar=|,
otherkeywords={table,var},
commentstyle=\color{olive}
}

\newtcblisting{plainlisting}[1][]{
  enhanced,
  top = 0.1pt, bottom=0.1pt,
  colframe=white,
  colback=white,
  fonttitle=\bfseries\scriptsize, 
  colbacktitle=white,
  coltitle=black,
  boxed title style={
    boxrule=0pt,
    colframe=white,
    },
  listing only,
  listing style=PythonStyle,
  notitle,
  #1}
  
\newtcblisting{simplelisting}[2][]{
  enhanced,
  attach boxed title to top right={yshift=-3mm,xshift=-2mm},
  top = 0.1pt, bottom=0.1pt,
  colframe=black,
  colback=white,
  fonttitle=\bfseries\scriptsize, 
  colbacktitle=white,
  coltitle=black,
  boxed title style={
    boxrule=0pt,
    colframe=white,
    },
  listing only,
  listing style=SimpleStyle,
  title=#2,
  #1}
  
\newtcblisting{hydrolisting}[2][]{
  enhanced,
  attach boxed title to top right={yshift=-3mm,xshift=-2mm},
  top = 0.1pt, bottom=0.1pt,
  colframe=black,
  colback=white,
  fonttitle=\bfseries\scriptsize, 
  colbacktitle=white,
  coltitle=black,
  boxed title style={
    boxrule=0pt,
    colframe=white,
    },
  listing only,
  listing style=HydroStyle,
  title=#2,
  #1}

\lstdefinestyle{SQLStyle} {
basicstyle=\ttfamily\scriptsize,       
language=SQL,
keywordstyle=\color{blue}\ttfamily\bfseries,
numbers=left,                   
numberstyle=\ttfamily\color{gray},      
stepnumber=1,                   
numbersep=8pt,                  
breakindent=0pt,
firstnumber=1,
showspaces=false,               
showstringspaces=false,         
showtabs=false,                 
frame=leftline,
tabsize=2,  		
captionpos=b,   		
breaklines=false,    	
breakatwhitespace=true,    
columns=fixed,
basewidth=0.52em,
numberblanklines=false,
escapechar=^,
morekeywords={next,async,declare,return,with,language,returns,function,handler,type},
alsoother={@}
}

  \newtcblisting{sqllisting}[2][]{
  enhanced,
  attach boxed title to top right={yshift=-3mm,xshift=-2mm},
  top = 0.1pt, bottom=0.1pt,
  colframe=black,
  colback=white,
  fonttitle=\bfseries\scriptsize, 
  colbacktitle=white,
  coltitle=black,
  boxed title style={
    boxrule=0pt,
    colframe=white,
    },
  listing only,
  listing style=SQLStyle,
  title=#2,
  #1}
  
\lstdefinestyle{BloomStyle} {
basicstyle=\ttfamily\scriptsize,       
language=ruby,
numbers=left,                   
numberstyle=\ttfamily\color{gray},      
stepnumber=1,                   
numbersep=8pt,                  
breakindent=0pt,
firstnumber=1,
showspaces=false,               
showstringspaces=false,         
showtabs=false,                 
frame=leftline,
tabsize=2,  		
captionpos=b,   		
breaklines=false,    	
breakatwhitespace=true,    
columns=fixed,
basewidth=0.52em,
numberblanklines=false,
escapechar=^,
otherkeywords={<+-},
keywordstyle=[1]{\color{blue}\ttfamily\bfseries},
keywordstyle=[2]{\color{red}\ttfamily\bfseries},
morekeywords=[1]{bloom,state,bootstrap,table,scratch,channel,interface,lset,lbool,lmax,lmap,lmin,in,out},
morekeywords=[2]{<+-,<+,<-}
}

  \newtcblisting{bloomlisting}[2][]{
  enhanced,
  attach boxed title to top right={yshift=-3mm,xshift=-2mm},
  top = 0.1pt, bottom=0.1pt,
  colframe=black,
  colback=white,
  fonttitle=\bfseries\scriptsize, 
  colbacktitle=white,
  coltitle=black,
  boxed title style={
    boxrule=0pt,
    colframe=white,
    },
  listing only,
  listing style=BloomLStyle,
  title=#2,
  #1}

\begin{document}

\title{New Directions in Cloud Programming}


\author{Alvin Cheung $\quad$ Natacha Crooks $\quad$ Joseph M. Hellerstein $\quad$  Matthew Milano}
\email{{akcheung,ncrooks,hellerstein,mpmilano}@berkeley.edu}
\affiliation{%
  \institution{UC Berkeley}
}


\begin{abstract}
Nearly twenty years after the launch of AWS, it remains difficult for most developers to 
harness the enormous potential 
of the cloud. 
In this paper we lay out an agenda for a new generation
of cloud programming research aimed at bringing research 
ideas to programmers in an evolutionary fashion. 
Key to our approach is a separation
of distributed programs into a PACT of four facets: Program semantics,
Availablity, Consistency and Targets of optimization. 
We propose to migrate developers gradually to PACT programming
by lifting familiar code into our more declarative level of abstraction.
We then propose a multi-stage compiler that emits human-readable code at
each stage that can be hand-tuned by developers seeking more control. 
Our agenda raises numerous research challenges across multiple areas including
language design, query optimization, transactions, distributed consistency, compilers and program synthesis.
\end{abstract}


\settopmatter{printacmref=false} 
\acmDOI{}
\acmISBN{}
\acmYear{1888}
\pagestyle{plain} 

\maketitle

\section{Introduction}
\label{sec:intro}
It is easy to take the public clouds for granted, but we have barely scratched the surface of their potential. These are the largest computing platforms ever assembled, and among the easiest to access. Prior generations of architectural revolutions led to programming models that unlocked their potential: minicomputers led to C and the UNIX shell, 
personal computers to graphical ``low-code'' programming via LabView and Hypercard, smartphones to Android and Swift. To date, the cloud has yet to inspire a programming environment that exposes the inherent power of the platform. 

Initial commercial efforts at a programmable cloud have started to take wing recently in the form of ``serverless'' Functions-as-a-Service. FaaS offerings allow developers to write sequential code and upload it to the cloud, where it is executed in an independent, replicated fashion at whatever scale of workload it attracts.  First-generation FaaS systems have well-documented limitations~\cite{hellerstein2019serverless}, which are being addressed by newer prototypes with more advanced FaaS designs (e.g., ~\cite{akkus2018sand,sreekanti2020cloudburst}). But fundamentally, even ``FaaS done right'' is a low-level assembly language for the cloud, a simple infrastructure for launching sequential code, a UDF framework without a programming model to host it.

As cloud programming matures, it seems inevitable that it will depart from traditional sequential programming. The cloud is a massive, globe-spanning distributed computer made up of heterogeneous multicore machines. Parallelism abounds at all scales, and the distributed systems challenges of non-deterministic network interleavings and partial failures exist at most of those scales. Creative programmers are held back by the need to account for these complexities using legacy sequential programming models originally designed for single-processor machines.

We need a programming environment that addresses these complexities directly, but without requiring programmers to radically change behavior.
The next generation of technology should \emph{evolutionize} the way developers program: allow them to address distributed concerns gradually, working with the assistance of new automation technologies, but retaining the ability to manually override automated decisions over time.  

\subsection{A New PACT for Cloud Programming}
\label{sec:acid30}

Moving forward, we envision decoupling cloud programming into four separate concerns, each with an independent language facet: Program semantics, Availability, Consistency and Targets for optimization (PACT).

\smallitem{{\Large P}rogram Semantics: Lift and Support.}
A programmer's primary goal is to specify the intended functionality of their program.
Few programmers can correctly write down their program semantics in a sequential language while also accounting for parallel interleavings, message reordering, partial failures and dynamically autoscaling deployment. This kind of ``hand-crafted'' distributed programming is akin to assembly language for the cloud.

Declarative specifications offer a very different solution, shielding the programmer from implementation and deployment details. 
Declarative programming environments for distributed computing have emerged in academia and industry over the past decade~\cite{alvaro2011consistency,datomic12,grangerlive2018}, but adoption of these ``revolutionary'' approaches has been limited. 
Moving forward, we 
advocate an evolutionary \emph{Lift and Support} approach: given a program specification written in a familiar style, automatically lift as much as possible to a higher-level declarative Intermediate Representation (IR) used by the compiler, and encapsulate what remains in UDFs (i.e., FaaS Functions).

\smallitem{{\Large A}vailability Specification.}
Availability is one of the key advantages of the cloud. Cloud vendors offer 
hardware and networking to deploy services redundantly across multiple relatively independent failure domains.
Traditionally, though, developers have had to craft custom solutions to ensure that their code and deployments take advantage of this redundancy efficiently and correctly. 
Availability protocols are frequently interleaved into program logic in ways that make them tricky to 
test and evolve.
We envision a declarative facet here as well, allowing programmers to specify the availability they wish to offer independent from their program semantics. A compiler stage must then synthesize code to provide that availability guarantee efficiently.

\smallitem{{\Large C}onsistency Guarantees.}
Many of the hardest challenges of distributed programming involve consistency guarantees.
``Sophisticated'' distributed programs are often salted with programmer-designed mechanisms to maintain consistency. We advocate a programming environment that separates consistency specifications into a first-class program facet, separated from the basic functionality. A compiler stage can then generate custom code to guarantee that clients see the desired consistency subject to availability guarantees. Disentangling consistency invariants from code makes two things explicit: the desired common-case sequential semantics,
and the relaxations of those semantics that are to be tolerated in the distributed setting.
This faceting makes it easier for compilers to guarantee correctness and achieve efficiency, 
it allows enforcement across compositions of multiple distributed libraries, 
and allows developers to easily understand and modify the consistency guarantees of their code. 


\smallitem{{\Large T}argets for Dynamic Optimization.}
In the modern cloud, code is not just compiled; it must be
deployed as a well-configured service across multiple machines. It also must be able to redeploy itself dynamically---\emph{autoscale}---to work efficiently as workloads grow and shrink by orders of magnitude, from a single multicore box to a datacenter to the globe.
We believe cloud frameworks inevitably must lighten this load for general-purpose developers. 
We envision an environment where programmers can specify multi-objective performance targets for execution, e.g., tradeoffs between billing costs, latency and availability. From there, a number of implementation and deployment decisions must be made. This includes compilation logic like choosing the right data structures and algorithms for ``local,'' sequential code fragments, as well as protocols for message-passing for distributed functionality. It also
includes the partitioning, replication and placement of code and data across machines with potentially heterogeneous resources. Finally, the binary executables we generate need to include dynamic runtime logic that monitors and adapts the deployment in the face of shifting workloads.
\smallitembot{}

For all these facets, we envision a \emph{gradual} approach to bring programmers on board in an evolutionary manner. Today's developers should be able to get initial success by writing simple familiar programs, and entrusting everything else to a compiler. In turn, this compiler should generate human-centric code in well-documented internal languages, suitable for eventual refinement by programmers.
As a start, we believe an initial compiler should be able to achieve performance and cost at the level of FaaS offerings that users tolerate today~\cite{hellerstein2019serverless}, with the full functionality of PACT programming.
Programmers can then improve the generated programs incrementally by modifying the lower-level facets or ``hinting'' the compiler via constraints. 

\subsection{Sources of Inspiration and Confidence}
Our goals for the next generation of cloud programming are ambitious, but work over the last decade gives us confidence that we can take significant strides in this direction.
A number of ideas from the past decade inform our approach: 

\smallitem{Monotonic Distributed Programming.}
Monotonicity---the property that a program's output grows with its input---
has emerged as a key foundation for efficient, available distributed programs~\cite{hellerstein2020keeping}. 
The roots of this idea go back to
Helland and Campbell's crystallization of coordination-free distributed design patterns as ACID 2.0: Associative, Commutative, Idempotent and Distributed~\cite{helland2009building}. Subsequently, CRDTs~\cite{shapiro2011conflict} were proposed as data types with ACI methods, observing that the ACI properties are those of join-semilattices: algebraic structures that grow monotonically. The connection between monotonicity and order-independence turns out to be fundamental.
The CALM Theorem~\cite{hellerstein2010declarative,ameloot2011relational} proved that programs produce deterministic outcomes without coordination \emph{if and only if} they are monotonic. 
Hence monotonic code can run coordination-free without any need for locking, barriers, commit, consensus, etc.
At the same time, our Bloom language~\cite{alvaro2011consistency,conway2012logic} adopted declarative logic programming for distributed computing,
with a focus on a monotonic core, and coordination only for non-monotone expressions. 
Various monotonic distributed language proposals have followed~\cite{kuper2013lvars,meiklejohn2015lasp,milano2019tour}. Monotonic design patterns have led to clean versions of complex distributed applications like collaborative editing~\cite{weiss2009logoot}, and high-performance, consistency-rich autoscaling systems like the Anna KVS~\cite{wu2019anna}. 

\smallitem{Dataflow and Reactive Programming.} 
Much of the code in a distributed application involves data that flows between machines, and event-handling at endpoints. Distributed dataflow is a notable success story in parallel and distributed computing,
from its roots in 1980s parallel databases~\cite{dewitt1992parallel} through to recent work on richer models like Timely Dataflow~\cite{mcsherry2017modular} and efforts to autoscale dataflow in the cloud~\cite{kalavri2018three}.
For event handling, reactive programming libraries like React.js~\cite{chedeau2014react} and Rx~\cite{meijer2010reactive} provide a different dataflow model for handling events and mutating state.
Given these successes and our experience with dataflow backends for low-latency settings~\cite{loo2009declarative,loo2005implementing,alvaro2010boom}
we are optimistic that a combination of dataflow and reactivity would provide a good general-purpose runtime target for services and protocols in the cloud. We are also encouraged by the general popularity of libraries like Spark and React.js---evidence that advanced programmers will be willing to customize low-level IR code in that style. 


\smallitem{Faceted Languages.} 
The success of LLVM~\cite{lattner2004llvm} has popularized the idea of multi-stage compilation with explicit internal representation (IR) languages.
We are inspired by the success of faceted languages and separation of concerns in systems design, with examples such as the model-view-controller design pattern for building user interfaces~\cite{design-patterns}, the three-tier architecture for web applications~\cite{ror, django}, and domain-specific languages such as Halide for image processing pipelines~\cite{ragan2013halide}.  
Dissecting an application into facets enables the compiler designer and runtime developer to choose different algorithms to translate and execute different parts of the application. For instance, 
Halide decouples algorithm specification from execution strategy, but keeps both as syntactic constructs for either programmer control or compiler autotuning. This decoupling has led Halide to outperform expert hand-tuned code that took far longer to develop, and its outputs are now used in commercial image processing software. Image processing is particularly inspiring, given its requirements for highly optimized code including  parallelism and locality in combinations of CPUs and GPUs.

\smallitem{Verified Lifting.}
Program synthesis is one of the most influential and promising practical breakthroughs in modern programming systems research~\cite{synthesisSurvey}.
Verified lifting is a technique we developed that uses program synthesis to formulate code translation as {\em code search}.
We have applied verified lifting to translate code across different domains, e.g., translating imperative Java to declarative SQL~\cite{qbs} and functional Spark~\cite{casper, casper2},  translating imperative C to CUDA kernels to Halide~\cite{dexter} and to hardware description languages~\cite{domino}. Our translated Halide code is now shipping in commercial products. Verified Lifting cannot handle arbitrary sequential code, but our Lift and Support approach should allow us to use it as a powerful programmer aid.

\smallitem{Client-Centric and Mixed Consistency.} 
Within the enormous literature on consistency and isolation, two recent thrusts 
are of particular note here.
Our recent work on \emph{client-centric} consistency steps away from traditional low-level histories to offer guarantees about what could be observed by a calling client. This has led to a new understanding of the connections between transactional isolation and distributed consistency guarantees~\cite{crooks2020client}.
Another theme across a number of our recent results is the \emph{composition} of services that offer different consistency guarantees~\cite{milano2019tour,milano2018mixt,crooks2016tardis}. 
The composition of multiple services with different consistency guarantees is a signature of modern cloud computing that needs to be brought more explicitly into programming frameworks. 
\smallitembot{}

\subsection{Outline}
In this paper we elaborate on our vision for cloud-centric programming technologies. We are exploring our ideas by building a new language stack called \sys that we introduce next. 
The development of \sys is part of our methodology, but we believe that the problems we are addressing can inform other efforts towards a more programmable cloud.

In Section~\ref{sec:overview} we provide a high-level overview of the \sys stack and a
scenario that we use as a running example in the paper. In Section~\ref{sec:logic} we present our ideas for \logic's program semantics facet, and in Section~\ref{sec:lifting-intro} we back up and explore the challenge of lifting from multiple distributed programming paradigms into \logic. Section~\ref{sec:data-model-facet} discusses the data model of \logic and our ability to use program synthesis to automatically choose data representations to meet performance goals. Section~\ref{sec:availability} sketches our first explicitly distributed language facet: control over Availability, while Section~\ref{sec:consistency} covers the challenges in the correlated facet of Consistency. 
In Section~\ref{sec:flow} we discuss lowering \logic to the corresponding \runtime algebra on a single node.
Finally, in Section~\ref{sec:slo} we address the distributed aspects of optimizing and deploying \runtime, subject to multi-objective goals for cost and performance.

\section{Hydro's Languages}
\label{sec:overview}
\begin{figure}
    \centering
    \includegraphics[width=\linewidth]{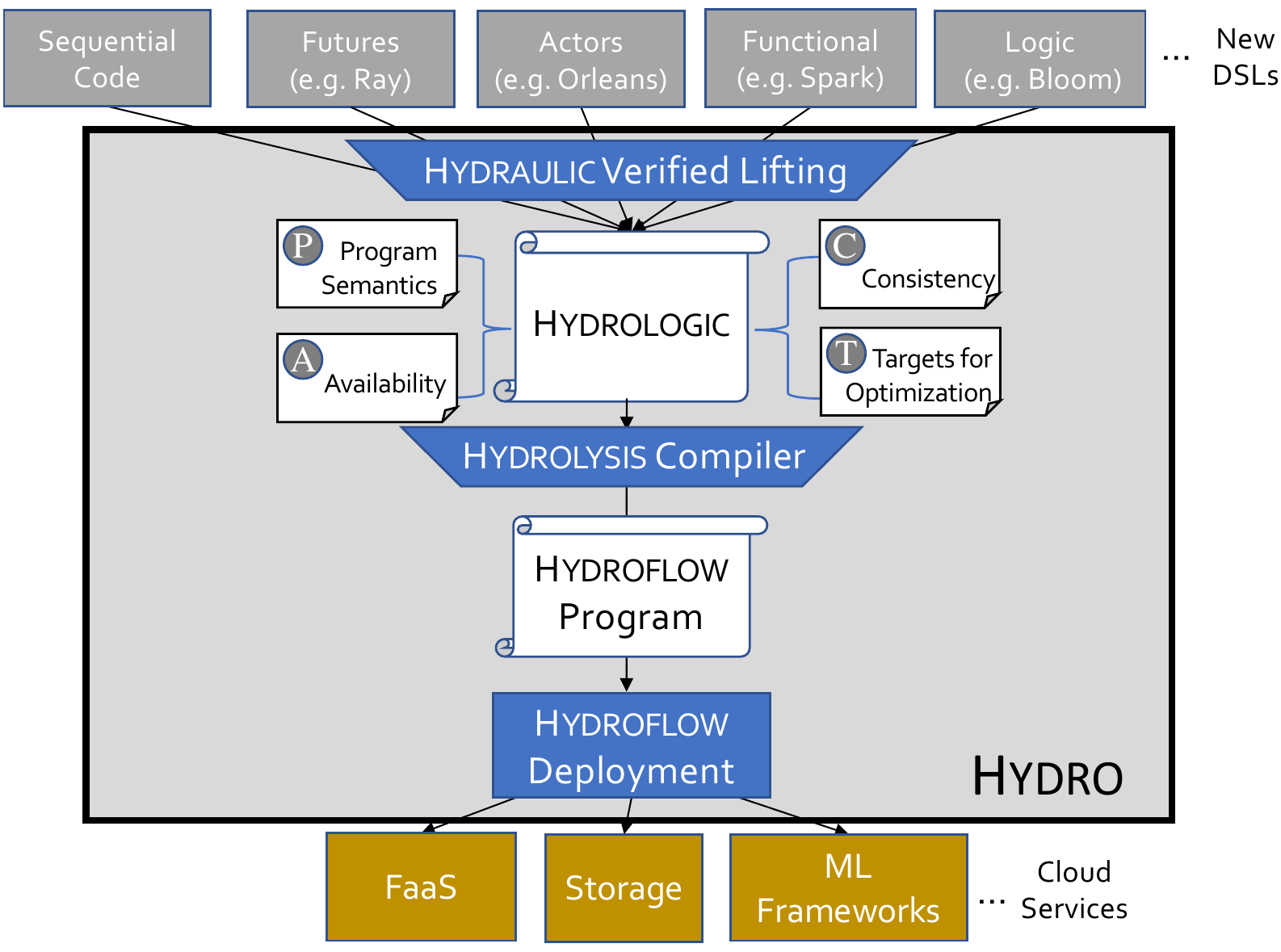}
    \caption{\rm\small The \sys stack.}
    \vspace{-0.2in}
    \label{fig:arch}
\end{figure}

\hydro consists of a faceted, three-stage compiler that takes programs in one or more distributed DSLs and compiles them to run on a low-level, autoscaling distributed deployment of local programs in the \runtime runtime. 

To bring a distributed library or DSL to the \hydro platform, we need to lift it to \hydro's declarative Intermediate Representation (IR) language---\logic. 
Hence our first stage is the \lifter Verified Lifting facility (Section~\ref{sec:lifting-intro}), which automates that lifting as much as it can, encapsulating whatever logic remains in UDFs.

\subsection{The Declarative Layer: \logic}
The \logic IR (Section~\ref{sec:logic}) is itself multifaceted as seen in Figure~\ref{fig:arch}. The core of the IR allows \emph{Program Semantics} \circledgray{P} to be captured in a declarative fashion, without recourse to implementation details regarding deployment or other physical optimizations.
For fragments of program logic that fail to lift to \logic, the language supports legacy sequential code via UDFs executed inline or asynchronously in a FaaS style.
The \emph{Availability} facet \circledgray{A} allows a programmer to ensure that each network endpoint in the application can remain available in the face of $f$ failures across specified failure domains (VMs, data centers, availability zones, etc.)
%
In the \emph{Consistency} facet \circledgray{C} we allow users to specify their desires for replica consistency and transactional properties; we use the term ``consistency'' to cover both. Specifically, we allow service endpoints to specify the consistency semantics that senders can expect to see.
The final high-level \emph{Targets for Optimization} facet \circledgray{T} allows the developer to specify multiple objectives for performance, including latency distributions, billing costs, downtime tolerance, etc.

Given a specification of the above facets, we have enough information to compile an executable cloud deployment. \logic's first three facets identify a finite space of satisfying distributed programs, and the fourth provides performance objectives for optimization in that space.

\subsection{Compilation of \logic}

The next challenge is to compile a \logic specification into an executable program deployable in the cloud. Rather than generating binaries to be deployed directly on different cloud platforms, we will instead compile \logic specifications into programs written against APIs exposed by the \runtime runtime (to be discussed in Section~\ref{sec:flow-overview}). Doing so allows experienced developers to fine-tune different aspects of a deployment while simplifying code generation. We are currently designing the \runtime APIs; we envision them to cover different primitives that can be used to implement the \logic facets, such as:
\begin{itemize}
    \item The choice of data structures for collection types and concrete physical implementations (e.g., join algorithm) to implement the semantics facet running as a local data flow on a single node.

    \item Partitioning (``sharding'') strategies for data and flows among multiple nodes, based on the data model facet.
    
    \item Mechanisms that together form the isolation and replica consistency protocols specific to the application.
    
    \item Scheduling and coordination primitives to execute data flows across multiple nodes, such as spawning and terminating \runtime threads on VMs.
    
    \item Monitoring hooks inserted into each local data flow to trigger adaptive reoptimization as needed during execution.
\end{itemize}


These primitives cover a lot of design ground, and we are still exploring their design. A natural initial approach is to provide a finite set of choices as different API calls, and combine API calls into libraries that provide similar functionalities for the compiler or developer to invoke (e.g., different data partitioning mechanisms). We imagine that the \compiler compiler will analyze multiple facets to determine which APIs to invoke for a given application, for instance combining the program semantics and targets for optimization facets to determine which data structures and physical implementations to use.
In subsequent sections we illustrate one or more choices for each. Readers familiar with these areas will hopefully begin to see the larger optimization space we envision, by substituting in prior work in databases, dataflow systems, and distributed storage. Note also that many of these features can be bootstrapped in \logic, e.g., by adapting prior work on distributed logic for transaction and consensus protocols~\cite{alvaro2010boom,alvaro2010declare}, query compilation~\cite{condie2008evita}, and distributed data structures~\cite{loo2005implementing}.

We are designing a compiler called \compiler to take a \logic specification and generate programs to be executed on the \runtime runtime.
As mentioned, our initial goal for \compiler is to guarantee correctness while meeting the performance of deployments on commercial FaaS pipelines. Our next goal is to explore different compilation strategies for \compiler, ranging from syntax-directed, cost-driven translation (similar to a typical SQL optimizer), to utilizing program synthesis and machine learning for compilation. The faceted design of \logic makes it easy to explore this space: each facet can be compiled independently using (a combination of) different strategies, and the generated code can then be combined and further optimized with low-level, whole program transformation passes. 




\subsection{The Executable Layer: \runtime}
\label{sec:flow-overview}
The \runtime runtime (Section~\ref{sec:flow}) is a strongly-typed single-node flow runtime implemented in Rust. It subsumes ideas from both the dataflow engines common in data processing, and the reactive programming engines more commonly used in event-driven UI programming. \runtime provides an event-driven, flow-based execution model, with operators that produce and consume various types including collections (sets, relations, tensors, etc.), lattices (counters, vector clocks, etc.) and traditional mutable scalar variables. 

\runtime executes within a transducer network~\cite{ameloot2011relational} (Section~\ref{sec:semantics}).
This event model allows for very high efficiency: as in the high-performance Anna KVS~\cite{wu2019anna}, all state is thread local and \runtime does not require any locks, atomics, or other coordination for its own execution.  Another advantage of the transducer model is the clean temporal semantics. As discussed in Section~\ref{sec:semantics}, all state updates are deferred to end-of-tick and applied atomically, so that handlers do not experience race conditions within a tick. Non-deterministic ordering arises only via explicit asynchronous messages.

\subsection{A Running Example}
As a running example, we start with a simplified backend for a COVID-19 tracking app. We assume a front-end application that generates pairwise contact traces, allows medical organizations to report positive diagnoses, and alerts users to the risk of infection. Sequential 
pseudocode is in Figure~\ref{fig:tracker}. 

\begin{figure}
    \input{ex_pseudo}
    \caption{\rm\small A simple COVID-19 tracking application.}
    \label{fig:tracker}
\end{figure}
The application logic starts with basic code to add an entry to the set \texttt{people}. 
The \texttt{add\_contact} function records the arrival of a new contact pair in the \texttt{contacts} list of both \texttt{people} involved. 
The utility function \texttt{trace} returns the transitive closure of a person's contacts. Upon diagnosis, the \texttt{diagnosed} function updates the
state and sends an alert to the app for every person transitively in contact. 
Next up is the \texttt{likelihood} function, which allows recipients of an alert to
synchronously invoke an imported black-box ML model \texttt{covid\_predict}, which returns a likelihood that the virus propagated to them through the contact graph. 

Our final function allocates a vaccine from inventory to a particular person.
We will revisit this example shortly, lifted into \logic.




\section{The Program Semantics Facet}
\label{sec:logic}
In \sys, our ``evolutionary'' approach is to accept programs written in sequential code or
legacy distributed frameworks like actors and futures. In a best-effort fashion, we lift these programs into a higher-level Internal Representation (IR) language called \logic. 
Over time we envision
a desire among some programmers for a more ``revolutionary'' approach involving user-friendly syntax that maps fairly directly to \logic or \runtime and their more optimizable constructs. The IR syntax we present here is preliminary and designed for exposition; we leave the full design of \logic syntax for future work.

We want our IR to be a target that is \emph{optimizable}, \emph{general} and \emph{programmer-friendly}. In the next few sections we introduce the IR and the ways in which 
it is amenable to distributed optimizations. In Appendix~\ref{sec:lifting-appendix} we demonstrate generality by showing how various distributed computing models can compile to \logic.


Figure ~\ref{fig:hydrotracker} shows our running example in a Pythonic version of \logic. 
The data model is presented in lines~\ref{line:datamodel} through~\ref{line:vaccine_count}, discussed further in Section~\ref{sec:data-model-facet}.
The program semantics (Section~\ref{sec:semantics}) are specified in lines~\ref{line:startofapp} through \ref{line:endofapp}, with the consistency facet (Section~\ref{sec:consistency}) declared inline 
for the handler at Line~\ref{line:vaccinate} that does not use the default of \texttt{eventual}. Availability (Section~\ref{sec:availability}) and Target facets (Section~\ref{sec:slo}) appear at the end of the example.

\subsection{\logic Semantics}
\label{sec:semantics}

\begin{figure}
    \input{ex_hydro}
    \caption{\rm\small A simple COVID-19 tracking application in a Pythonic \logic syntax.
    Each \texttt{\ttfamily\bfseries on} handler has faceted specifications of consistency, availability and deployment either in the definition (as is done here with consistency specs) or defined in a separate block.}
    \vspace{-0.1in}
    \label{fig:hydrotracker}
\end{figure}

\logic's program semantics begin with its event loop, which is based on the transducer model in Bloom~\cite{alvaro2011consistency}. 
\logic's event loop considers the current \emph{snapshot} of program state, which includes any new inbound messages to be handled. Each iteration (``tick'') of the loop uses the developer's program specification to compute new results from the snapshot, and atomically updates state at the end of the tick. All computation within the tick is done to fixpoint. The snapshot and fixpoint semantics together ensure that the results of a tick are independent of the order in which statements appear in the program.

The notion of endpoints and events should be familiar to developers of microservices or actors. Unlike microservices, actors or Bloom, \logic's application semantics provide a simple ``single-node'' model---a global view of state, and a single event loop providing a single sequence (clock) of iterations (ticks). 
This single-node metaphor is part of the facet's declarative nature---it ignores issues of data placement, replication, message passing, distributed time and consistency, deferring them to separable facets of the stack.

Basic statements in \logic's program semantics facet come in a few forms:

\smallitemem{Queries} derive information from the current snapshot. Queries are named and referenceable, like SQL views, and defined over various lattice types,
including relational tables. Line~\ref{line:transitive1} represents a simple query returning pairs of {\tt Person}s, the second of whom is a contact in the first.
As in Datalog, multiple queries can have the same name, implicitly defining a merge of results across them. Lines~\ref{line:transitive1} and~\ref{line:transitive2} are an example, defining the base case and inductive case, respectively, for graph transitive closure\footnote{\logic supports
recursion and non-monotonic operations (with stratified negation) for both relations and lattices. These features are based on Bloom$^L$ and the interested reader is referred to~\cite{conway2012logic} for details.}. 
A query $q$ may have the same name as a data variable $q$, in which case the contents of data variable $q$ are implicitly included in the query result.

\smallitemem{Mutations} are requests to modify data variables based on the current contents of the snapshot. 
Following the transducer model, mutations are deferred until the end of a clock ``tick''---they become visible together, atomically, once the tick completes. 
Mutations take three forms. A lattice merge mutation as in lines~\ref{line:addperson1},\ref{line:addcontact1},\ref{line:addcontact2}, or \ref{line:vaccinate1} monotonically ``merges in'' the lattice value of its argument. The traditional bare assignment operator {\color{red}\texttt{:=}}, as in line~\ref{line:vaccinate2} represents an arbitrary, likely non-monotonic update. 
A query $q$ with the same name as a data variable $q$ implicitly replaces (mutates) $q$ at end of tick; this mutation is monotonic iff the query is monotonic.

\smallitemem{Handlers} begin with the keyword {\ttfamily\bfseries on}, and model reactions to messages.
Seen within the confines of a tick, though, a handler is simply syntactic sugar for \hydro statements mapped over a \emph{mailbox} of messages corresponding to the handler's name. 
The body of a handler is a collection of \logic statements, each quantified by the particular message being mapped.
For example, the \texttt{add\_person} handler on Line~\ref{line:addperson} is syntactic sugar for the \logic statements: 
\begin{lstlisting}[style=HydroStyle]
people.merge(Person(a.pid) for a in add_person)
send add_person<response>(message_id: int, payload: Status):
  {(a.message_id, OK) for a in add_person}
\end{lstlisting}
The implicit mailbox \texttt{add\_person<response>} is used to send results to the caller of an \texttt{add\_purpose} API---e.g., to send the HTTP status response to a REST call.

\smallitemem{UDFs} are black-box functions, and may keep internal state across invocations. An example UDF, \texttt{covid\_predict}, can be seen in the \texttt{likelihood} handler of line~\ref{line:likelihood}. UDFs cannot access \logic variables and should avoid any other external, globally-visible data storage. Because UDFs can be stateful and non-idempotent, each UDF is invoked once per input per tick (memoized by the runtime), in arbitrary order.

\smallitemem{Send} is an asynchronous merge into a mailbox.
As with mutations, sends are not visible during the current tick. Unlike mutations, sends might not appear atomically---each individual object sent from a given tick may be ``delayed'' an unbounded number of ticks, appearing non-deterministically in the specified mailbox at any later tick. Sends capture the semantics of unbounded network delay. Line~\ref{line:diagnosed2} provides an internal example, letting the compiler know that we expect alerts to be delivered asynchronously.
As another example, we can rewrite the \texttt{likelihood} handler of line~\ref{line:likelihood} to use a remote FaaS service. This requires sending a request to the service and handling a response:
\begin{lstlisting}[style=HydroStyle]
on async_likelihood(pid:int, isolation=snapshot)
  send FaaS((covid_predict, handler.message_id, find_person(pid)))

on covid_predict<response>(al_message_id: int, result: bool):
  send async_likelihood<response>((handler.message_id, 
                                  al_message_id, result))
\end{lstlisting}

\smallitemembot

\logic statements can be bundled into \emph{blocks} of multiple statements, as in the bodies of the \texttt{add\_contact} and \texttt{vaccinate} handlers. Blocks can be declared as object-like \emph{modules} with methods to scope naming and allow reuse. Blocks and modules are purely syntactic sugar and we do not describe them further here.

\section{Lifting to \logic}
\label{sec:lifting-intro}
We aim for \logic to be an evolutionary, general-purpose IR that can be targeted from a range of 
legacy design patterns and languages, while pointing the way toward coding styles that take advantage of more recent research.

Our goal in the near term is not to convert any arbitrary piece of code into an elegant, easily-optimized \logic program. In particular, we do not focus on lifting existing ``hand-crafted'' distributed programs to \logic. We have a fair bit of experience (and humility!) about such a general goal. Instead we focus on two scenarios for lifting:

\smallitem{Lifting single-threaded applications to the cloud:}
Many applications consist largely of single-threaded logic, but would benefit from scaling---and autoscaling---in the cloud.
In our earlier work, we have had success using verified lifting to convert sequential imperative code of this sort into declarative frameworks like SQL~\cite{qbs}, Spark~\cite{casper, casper2} and Halide~\cite{dexter}. 
One advantage of sequential programs---as opposed to hand-coded multi-threaded or distributed ``assembly code''---is that we do not have to reverse-engineer consistency semantics from ad hoc patterns of messaging or concurrency control in shared memory. 
Some interesting corpora of applications are already written in opinionated frameworks that assist our goals. For example, applications that are built on top of object-relational mapping (ORM) libraries such as Rails~\cite{ror} and Django~\cite{django} are essentially built on top of data definition languages (e.g., ActiveRecord~\cite{activerecord}), which makes it easy to lift the data model, and sometimes explicit transactional semantics as well. ORM-based applications also often serve as backends for multiple clients and need to scale over time---Twitter is a notorious example of a Rails app that had to be rewritten for scalability and availability.

\smallitem{Evolving a breadth of distributed programming frameworks:} There are existing distributed programming frameworks that are fairly popular, and our near-term goal is to embrace these programming styles. Simple examples include FaaS interfaces and big-data style functional dataflow like Spark. Other popular examples for asynchronous distributed systems include actor libraries (e.g., Erlang~\cite{erlang}, Akka~\cite{akka}, Orleans~\cite{bykov2011orleans}), libraries for distributed promises/futures (e.g., Ray~\cite{moritz2018ray} and Dask~\cite{dask} for Python), and collective communication libraries like that of MPI~\cite{mpi-collective}. Programs written with these libraries adhere to fairly stylized uses of distributed state and computation, which we believe we can lift relatively cleanly to \logic. In Appendix~\ref{sec:lifting-appendix} we share our initial thoughts and examples in this direction.
\smallitembot

Our goals for lifting also offer validation baselines for the rest of our research. If we can lift code from popular frameworks, we can auto-generate a corpus of test cases. \sys should aim to compete with the native runtimes for these test cases. In addition, lifting to \logic will hopefully illustrate the additional flexibility \sys offers via faceted re-specification of consistency, availability and performance goals. And finally, success here across different styles of frameworks will demonstrate the viability of our stack as a common cloud runtime for multiple styles of distributed programming, old and new.




\section{\logic's Data Modeling}
\label{sec:data-model-facet}
\logic data models consist of four components: 1) a class hierarchy that describes how persistent data is structured, 2) relational constraints, such as functional dependencies, 3) persistent collection abstractions like relations, ordered lists, sets, and associative arrays, and 4) declarations for data placement across nodes in distributed deployments. 

For instance, Lines~\ref{line:datamodel}-\ref{line:vaccine_count} in Figure~\ref{fig:hydrotracker} show an example of persistent data specification for our Covid application. The data is structured as {\tt Person} objects, each storing an integer {\tt pid} that serves as a unique id (key), along with a set of references to other {\tt Person}s that they have been in contact with. 
Line~\ref{line:partition} illustrates an optional {\tt partition}
value to suggest how {\tt Person} objects should be partitioned across multiple nodes. (\logic uses the class's unique id to partition by default).
Line~\ref{line:people} then prescribes that the {\tt Person}s are to be collectively stored in a {\tt table} keyed on each person's {\tt pid} that is publicly accessible by all functions in the program. 

Partitioning allows developers to hint at ways to scatter data; a similar syntax for locality hints is available. These hints are not required, however:
\logic programmers can define their data model without needing to know how their data will be stored in the cloud. The goal of \sys is to take such user-provided specifications and generate a concrete implementation afterwards.

\subsection{Design Space}

As part of compilation, we need to choose an implementation of the data model facet. For example in our Covid tracker we might store {\tt Person} objects in memory using an associative array indexed on each person's {\tt pid}, with each person's {\tt contacts} field stored as a list with only the {\tt pid}s of the {\tt Person} objects. 
Obviously this particular implementation choice has tradeoffs with more normalized choices, depending on workload.



In general, a concrete data structure implementation consists of two components: choosing the container(s) to store persistent data (e.g., a B+-tree indexed on a field declared in one of the persistent classes), and determining the access path(s) given the choices for containers (e.g., an index or full container scan) when looking up a specific object.

We envision that there will be multiple algorithms to generate concrete implementations. These can range from a rule-driven approach that directly matches on specific forms of queries and determines the corresponding implementation (e.g., for programs with many lookup queries based on {\tt id}, use an associative array to store {\tt Person} objects), to a synthesis-driven approach that enumerates different implementations based on a grammar of basic data structure implementations~\cite{stratos-data-structures} and a cost model. Access paths can then be determined based on how the containers are selected. 

\subsection{Promise and Challenges}
\label{sec:datamodelpromise}
We have designed a data structure synthesizer called Chestnut in our earlier work~\cite{chestnut, chestnut-demo}, focusing on database-backed web applications that are built using ORM libraries. Similar to \logic, Chestnut takes in a user-provided data model and workload specification, and synthesizes data structures to store persistent data once it is loaded into memory. Synthesis is done using an enumeration-based approach based on a set of provided primitives. For example, if a persistently stored class contains N attributes, Chestnut would consider storing all objects in an ordered list, or in an associative array keyed on any of the N unique attributes, or split the N attributes into a subset that is stored in a list, and the rest stored in a B+-tree index. The corresponding access path is generated for each implementation. Choosing among the different options is guided by a cost model that estimates the cost of each query that can potentially be issued by the application. Evaluations using open-source web apps showed that Chestnut can improve query execution by up to 42$\times$.

Searching for the optimal data representation is reminiscent of the physical design problem in data management research, and there has been a long line of work on that front~\cite{autoadmin} that we can leverage.
There has also been work done on data structure synthesis in the programming systems research community~\cite{cozy, aiken-data-structures} that focuses on the single-node setting, with the goal to organize program state as relations and persistently store them as such.

Synthesizing data structures based on \logic specifications will raise new challenges. First, we will need to design a data structure programming interface that is expressive enough for the program specifications that users will write. Next, we will need a set of data structure ``building blocks'' that the synthesizer can utilize to implement program specifications. Such building blocks must be composable such that new structures can be designed, yet not too low-level that makes it difficult to verify if the synthesized implementation satisfies the provided specifications.

In addition, synthesizing distributed data structures will require new innovations in devising cost models for data transfer and storage costs, and reasoning about data placement and lookup mechanisms. New synthesis and verification algorithms will need to be devised in order to handle both aspects efficiently. Finally, workload changes (both client request rates and cloud service pricing) motivate incremental synthesis, where initial data structures are generated when the program is deployed, and gradually refined or converted to other implementations based on runtime properties.

\section{The Availability Facet}
\label{sec:availability}



The availability facet starts with a simple programmer contract: ensure that each application endpoint remains available in the face of $f$ independent failures. 
In this discussion we assume that failures are non-Byzantine. 
The definition of independence here is tied to a user-selected notion of \emph{failure domains}: two failures are considered independent if they are in different failure domains. Typical choices for failure domains include virtual machines, racks, data centers, or availability zones (AZs). 
In line~\ref{line:availability1} of Figure~\ref{fig:hydrotracker} we specify that our handlers should tolerate faults across 2 AZs. In line~\ref{line:availability2} we override that spec for the case of the \texttt{likelihood} handler, an ML routine that requires expensive GPU reservations, for which we trade off availability to save cost.


\subsection{Design Space}
The natural methodology for availability is to replicate service endpoints---execution and state---across failure domains. This goes back to the idea of process pairs in the Tandem computing systems, followed by the Borg and Lamport notions of state machine replication, and many distributed systems built thereafter. For example, when compiling the handler for the \texttt{add\_contact} endpoint in line~\ref{line:addcontact} of Figure~\ref{fig:hydrotracker}, we can interpose \logic implementing a load-balancing client proxy module that tracks replicas of the endpoint, forwards requests on to $f+1$ of them, and makes sure that a response gets to the client. 

  
Another standard approach is for backend logic to replicate its internal state---often by generating logs or lineage of state mutation events for subsequent replay. We could do this naively in our example by matching each mutation statement with a log statement.


\subsection{Promise and Challenges}
Whether using replication or replay, availability is fundamentally achieved by \emph{redundancy} of state and computation.
The design of that redundancy is typically complicated by two issues. The first is cost. In the absence of failure, redundancy logic can increase latency. Worse, running an identical replica of a massive service could be massively expensive. As a result, some replication schemes reduce the cost of replicas by having them perform logic that is different from---but semantically equivalent to---state change at the main service. 
A standard example is to do logical logging at the storage level, without redundantly performing application behavior.
In general, it would of course be challenging to synthesize sophisticated database logging and recovery protocols from scratch.
But simpler uses of activity logs for state replication are an increasingly common design pattern for distributed architectures~\cite{kafka,corfu,ambrosia}, and use of these basic log-shipping patterns and services could offer a point in the optimization space of latency, throughput and resource consumption that differs from application-level redundancy.


The second complication that arises immediately from availability is the issue of consistency across redundant state and computation, which we address next with its own facet.


\section{The Consistency Facet}
\label{sec:consistency}
The majority of distributed systems work has relegated consistency issues to the storage or memory layer. 
But the past decade has seen a variety of clever applications (shopping carts~\cite{decandia2007dynamo}, collaborative editing systems~\cite{weiss2009logoot}, gradiant descent~\cite{recht2011hogwild}, etc.) that have demonstrated massive performance and availability benefits by customizing consistency at the application layer. In \sys we aim to take full programs---including compositions of multiple independent modules---and automatically generate similarly clever, ``just right'' code to meet \emph{application-specific} consistency specifications.

The idea of raising transactional consistency from storage to the programming language level is familiar from object databases~\cite{atkinson1990object} and distributed object systems. Liskov's Argus language~\cite{liskov88argus} is a canonical example, with each distributed method invoked as an isolated (nested) transaction, strictly enforced via locking and two-phase commit.
This provides strong correctness properties---unnecessarily strong, since
not every method call in a distributed application requires strong consistency or perfect isolation. From our perspective today, Argus and its peers passed up the biggest question they raised: if all the application code is available, how \emph{little} enforcement can the compiler use to provide those semantics? And what if those semantics are weaker than serializability? 

As seen in the example of Figure~\ref{fig:hydrotracker}, \logic allows consistency to be specified at the level of the client API handlers. Like all our facets, consistency can be specified inline with the handler definition (as in Figure~\ref{fig:hydrotracker}), or in a separate \texttt{consistency} block.
In practice, applications are built from code written by different parties for potentially different purposes. As a result the original consistency specs provided for different handlers may be heterogeneous within a single application. What matters in the end is to respect the (possibly heterogeneous) consistency that clients of the application can observe from its public interfaces. 


In Figure~\ref{fig:hydrotracker}, the \texttt{add\_person} handler uses default eventual consistency. This ensures that if the two \texttt{people} in the arguments are not physically co-located, then each person (and each replica) can be updated without waiting for any others. 

As a different example, the \texttt{vaccinate} handler specifies
\texttt{serializability} and a non-negative \texttt{vaccine\_count} constraint. 
We might be concerned that serializability for this handler will require strong consistency from other handlers. Close analysis shows this is not the case:
\texttt{vaccinate} is the only handler that references \texttt{vaccine\_count},
and all references to \texttt{people} are monotonic and hence reorderable---including the mutation in \texttt{vaccinate}. Hence if \texttt{vaccinate} completes for some \texttt{pid} in any history, there is an equivalent serial history in which \texttt{vaccinate(pid)} runs successfully with the same initial value of \texttt{vaccine\_count} and the same resulting value of both \texttt{vaccine\_count} and \texttt{people}.

\subsection{Design Space}
In \logic we enable two different types of consistency specifications: traditional history-based guarantees, and application-centric \emph{invariants}. History-based guarantees are prevalent today, with widely agreed-upon semantics. For example, serializability, linearizability, sequential consistency, causal consistency, and others specifically constrain the ordering of conflicting operations and in turn define ``anomalies'' that applications can observe. The second type of consistency annotation we allow is application-centric, and makes use of Hydrologic's declarative formulation. Past work has demonstrated that
invariants are a powerful way for developers to precisely specify what guarantees 
are necessary at application
level~\cite{hellerstein2020keeping,bailis2014coordination,whittaker2018interactive,magrino2019warranties,roy2015homeostatis,shapiro15invariant,sivaramakrishnan2015declarative,crooks2017seeing}.
These include motonicity invariants that guarantee convergent outcomes, or isolation invariants for predicates on visible states---e.g., positive bank accounts or referential integrity.

\subsection{Promise and Challenges}
Many challenges fall out from an agenda of compiling arbitrary distributed code to efficiently enforce consistency invariants. Based on work to date, we believe the field is ripe for innovation. Here we highlight some key challenges and our reasons for optimism. 

\smallitem{Metaconsistency Analysis}: Servicing a single public API call may require crossing multiple internal endpoints with different consistency specifications. This entails two challenges: identifying the possible composition paths, and ensuring \emph{metaconsistency}: the consistency of heterogeneous consistency specs along each path. The first problem amounts to dataflow analysis across \logic handlers; this is easy to do conservatively in a static analysis of a \logic program, though we may desire more nuanced conditional solutions enforced at runtime. The question of metaconsistency is related to our prior work on mixed consistency of black-box services~\cite{milano2019tour,milano2018mixt,crooks2016tardis}. In the \sys context we may use third-party services, but we also expect to have plenty of white-box \logic code, where we have the flexibility to change the consistency specs across modules to make them consistent with the consistency of endpoint specifications. Our recent work on client-centric consistency offers a unified framework for reasoning about both transactional isolation \emph{and} distributed consistency guarantees~\cite{crooks2020client}.

\smallitem{Consistency Mechanisms}: Given a consistency requirement, we need to synthesize code to enforce it. There are three broad approaches to choose from. The first is to recognize when no enforcement is required for a particular code block---examples include the monotonicity and invariant confluence analyses mentioned above. Another is for the compiler to wrap or ``encapsulate'' state with lattice metadata that allows for local (coordination-free) consistency enforcement at each endpoint---this is the approach in our work on the Cloudburst FaaS~\cite{sreekanti2020cloudburst} and Hydrocache~\cite{wu2020transactional}. The third approach is the traditional ``heavyweight'' use of coordination protocols, including barriers, transaction protocols, consensus-based logs for state-machine replication and so on. The space of enforcement mechanisms is wide, but there are well-known building blocks in the literature that we can use to start on our software synthesis agenda here.

\smallitem{Consistency Placement}: Understanding consistency specs and mechanisms is not enough---we can also reason about where to invoke the mechanism in the program, and how the spec is kept invariant downstream. This flexibility arises when we consider consistency at an application level rather than as a storage guarantee. As a canonical example, the original work on Dynamo's shopping carts was coordination-free \emph{except} for ``sealing'' the final cart contents for checkout \cite{decandia2007dynamo,helland2009building,alvaro2011consistency}. Conway~\cite{conway2012logic} shifted the sealing to the
end-user's browser code where it is decided unilaterally (for ``free'') in an unreplicated stage of the application. When shopping ends, the browser ships a compressed \emph{manifest} summarizing the final content of the cart. Maintaining the final cart state at the replicas then becomes coordination-free as well: each replica can eagerly move to checkout once its contents match the manifest. Alvaro systematized this sealing idea in Blazes~\cite{alvaro2014blazes}; more work is needed to address the variety of coordination guarantees we wish to enforce and maintain.

\smallitembot

Clearly these issues are correlated, so a compiler will have to explore the space defined by their combinations.

\section{The \runtime IR}
\label{sec:flow}

Like many declarative languages, to execute \logic we translate it down to a lower-level algebra of operators that can be executed in a flow style on a single node, or partitioned and pipelined across multiple nodes (Section~\ref{sec:slo}). Most of these operators are familiar from relational algebra and functional libraries like Spark and Pandas. Here we focus on the unique aspects of the \logic algebra.

\subsection{Design Space}
The \runtime algebra has to handle all the constructs of \logic's event loop.
One of the key goals of the \runtime algebra design is a \emph{unification of dataflow, lattices and reactive programming.} Typical runtimes implement a dataflow model of operators over streaming collections of individual items. This assumes that collection types and their operators are the primary types in any program. We want to accommodate lattices beyond collection types. For example, a \texttt{COUNT} query takes a set lattice as input and produces an integer lattice as output; we need the output of that query to ``pipeline'' in the same fashion as a set. In addition, to capture state mutation we want to adapt reactive programming models (e.g., React.js and Rx).
that provide ordered streams propagating changes to individual values over time. 

In deployment, a \sys program involves \runtime algebra fragments running at multiple nodes in a network, communicating via messages. Inbound messages appear at \runtime ingress operators, and outbound messages are produced by egress operators. These operators are agnostic to networking details like addressing and queueing, which are parameterized by the target facet. However, as a working model we can consider that a network egress point in \runtime can be parameterized to do explicit point-to-point networking, or a content-hash-based style of addressing. As a result, local \runtime algebra programs can participate as fragments of a wide range of deployment models, including parallel intra-operator partitioning (a la Exchange or MapReduce) as well as static dataflows across algebraic operators, or dynamic invocations of on-demand operators.
\smallitembot

\subsection{Promise and Challenges}
A program in \logic can be lowered (compiled) to a set of single-node \runtime algebra expressions in a straightforward fashion, much as one can compile SQL to relational algebra. 
Our concern at this stage of lowering is scoped to single-node, in-memory performance; issues of distributed computing are deferred to Section~\ref{sec:slo}.  The design space here is similar to that of traditional query optimization, and classical methods such as Cascades are a plausible approach~\cite{cascades}. We are also considering more recent results in program synthesis here, since they have shown promise in traditional query optimization~\cite{qbs,statusquo}.


The design of the \runtime algebra is a work in progress, and achieving a semantics that unifies all of its aspects is non-trivial. In addition, two other challenges arise naturally.

\begin{figure}[t]
    \centering
    \includegraphics[width=\linewidth]{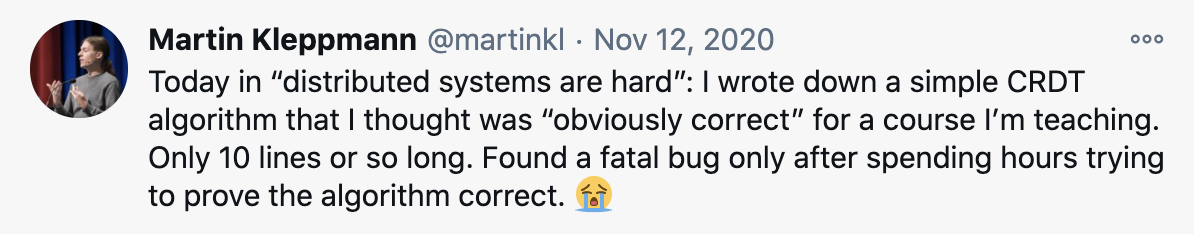}
    \caption{\rm On the trickiness of manual checks for monotonicity. The full thread includes pseudocode and fixes~\cite{kleppmanntweet}.}
    \vspace{-0.3in}
    \label{fig:kleppmann}
\end{figure}

\smallitem{Monotonicity typechecking:} 
Current models for monotonic programming like CRDT libraries expect programmers to guarantee the monotonicity of their code manually. This is notoriously tricky---see Figure~\ref{fig:kleppmann}. Bloom$^L$ attempted to simplify this problem by replacing monolithic CRDTs with monotone compositions of simpler lattices, but correctness was still assumed for the basic lattices and composition functions. We wish to go further, providing an explicit \texttt{monotone} type modifier, and a compiler that can typecheck monotonicity. Guarantees of monotonicity  can be exploited to ensure guarantees from the consistency facet (Section~\ref{sec:consistency}) as part of the Optimization facet (Section~\ref{sec:slo}).

\smallitem{Representation of flows beyond
collections:} 
Algebras defined for a collection type \texttt{C<T>} 
(e.g., relational algebra on \texttt{set<tuple>}) are often implemented 
in a dataflow of operators over the underlying element type \texttt{T},
or over incremental batches of elements. This \emph{differential} approach
is well-suited for operators on \texttt{C<T>} that have
stateless implementations over \texttt{T}---e.g., map, select and project. 
Other operators require stateful implementations that use ad-hoc internal memory management 
to rebuild collections of type \texttt{C<T>} across invocations over type \texttt{T}. This
makes it difficult for a compiler to check properties like determinism or
monotonicity. Moreover, in \runtime we want to expand flow computation beyond collection types to
lattices and reactive scalar values. 
Hence we need to support operators that view inputs differentially {\it or} all-at-once, providing clear semantics for both cases, and allowing
properties like monotonicity to be statically checked by a compiler.

\smallitem{Copy efficiency:} In many modern applications and systems, the majority of compute time is spent in copying and formatting data. Developers who have built high-performance query engines know that it is relatively easy to build a simple dataflow prototype, and quite hard to build one that makes efficient use of memory and minimizes the cost of data copying and redundant work. Taking a cue from recent systems like Timely Dataflow~\cite{mcsherry2017modular}, we use the ownership properties of the Rust language to help us carefully control how data is managed in memory along our flows.

\section{The Target Facet}
\label{sec:slo}


After specifying various semantic aspects of the application, the final facet describes the targets for optimization that the cloud runtime should achieve, as described in Section~\ref{sec:acid30}. Such targets can include a cost budget to spend on running the application on the cloud, maximum number of machines to utilize, specific capabilities of the hosted machines (e.g., GPU on board), latency requirements for any of the handlers, etc. We imagine that the user will provide a subset of these configuration parameters and leave the rest to be determined by \compiler.

For example, lines~\ref{line:deployment}-\ref{line:deploymentlikelihood} in Figure~\ref{fig:hydrotracker} show the targets for our COVID application. Line~\ref{line:deploymentdefault} specifies the default latency/cost goals for handlers; line~\ref{line:deploymentlikelihood} specializes this for machine-learning-based \texttt{likelihood} handler, dictating the use of GPU-class machines with a higher budget per call.

Compared to the current practice of deployment configurations spread across platform-specific scripts~\cite{aws-gateway, azure-management}, program annotations~\cite{cloud-annotations}, and service level agreements, \logic allows developers to consolidate deployment-related targets in an isolated program facet. This allows developers to easily see and change the cost/performance profiles of their code, and enables \logic applications to be deployed across different cloud platforms with different implementations.


\subsection{Design Space}



Given a \logic specification, the \compiler compiler will attempt to find an implementation that satisfies the provided constraints subject to the desired overall objectives. As discussed in~Section~\ref{sec:overview}, \sys will first generate an initial implementation of the application based on the previously described facets. The initial implementation would have various aspects of the application determined: algorithms for data-related operations, replication and consistency protocols, etc. What remains are the runtime deployment aspects such as mapping of functions and data to available machines. 

For instance, given the code in Figure~\ref{fig:hydrotracker}, \sys can formulate the runtime mapping problem as an integer programming problem, based on our prior work~\cite{pyxis, quro}. Such a mapping problem can be formulated as a dynamic program partitioning problem. Suppose at any given time we have $M$ different types of machine configurations to choose from, and $n_i$ represents the number of instances we will use for machine of type $i$. We then have the following constraints:

\begin{adjustwidth}{0.75em}{}
\smallitem{$\sbullet[.9]$} $latency( \texttt{add\_person}, n_i ) \le 100ms$. The latency incurred by hosting {\tt add\_person} on $n_i$ instances of type $i$ machines must be less than the specified value. We have one constraint for each pair of handler and machine type, and Figure~\ref{fig:hydrotracker} shows a shortcut using the {\tt default} construct while overriding it for the {\tt likelihood} handler.


\smallitem{$\sbullet[.9]$} $cost(\texttt{add\_person}, n_i) \le 0.01$. The cost of running {\tt add\_person} on $n_i$ instances of type $i$ machines must be less than the specified value. The value can either be specified by the end user or provided by the hosting platform.

\smallitem{$\sbullet[.9]$} $\sum_i n_i > 0$. Allocate some machines to fulfill the workload.
\smallitembot
\end{adjustwidth}
\vspace{-1em}

The overall objective depends on the user specification, for instance minimizing 
the total number of machines used ($\sum_i n_i$), or maximizing overall throughput of each handler $f_j$ while executed on $n_i$ machines ($\sum_{i,j} tput(f_j, n_i)$). More sophisticated objectives are possible, for instance incurring up to a fixed cost over a time period~\cite{aws-budgets}.

As formulated above, our integer programming problem relies on having models to estimate latency, throughput, and cost of running each function given machine type and number of instances. Solving the problem gives us the values of each $n_i$, i.e., the number of instances to allocate for each machine configuration. 
Given that program objectives or resource costs 
can vary as the application executes in the cloud, we might need to periodically reformulate the problem based on the data available. Predicting or detecting when a reformulation is needed will be interesting future work.

Note that the above integer program might not have a solution, 
e.g., if the initial implementations were too costly to meet the given targets. If this arises, \sys can ask previous components to choose other implementations and reiterate the mapping procedure. This iterative process is simplified by decomposing the application into facets, allowing \sys to revert to a previous search state during compilation.

\subsection{Promise and Challenges}
Our problem formulation above is inspired by prior work in multi-query optimization~\cite{mqo}, multi-objective optimization~\cite{multi-objective-qo} and scheduling tasks for serverless platforms~\cite{serverless-scheduling}. 
Our faceted setting, however, also raises new challenges.\\

\vspace{-1em}
\noindent {\bf Cost modeling:}
Our integer programming problem formulation relies on having accurate cost models for different aspects of program execution on the cloud (e.g., latency prediction). While cost prediction has been a classic research topic in data management, much of the prior work has focused on single and distributed relational databases. The cloud presents new challenges as functions can move across heterogeneous machines, and aspects such as machine prices and network latencies can vary at any time.

\noindent {\bf Solution enumeration:}
As mentioned earlier, our faceted approach allows \sys to easily backtrack during compilation, should an initial strategy turn out to be infeasible given the configuration constraints. Implementing backtracking will rely on an efficient way to enumerate different implementations based on the previously described facets, and being able to do so efficiently in real time. This depends on the algorithms used to generate the initial implementations, for instance by considering types of query plans that were previously bypassed during code generation, or asking solvers to generate another satisfiable solution if formal methods-based algorithms are used. We will also need feedback mechanisms to interact with the user, should the provided specifications prove too stringent.

\noindent {\bf Adaptive optimization:}
One of the reasons for deploying applications on the cloud is to leverage the cloud's elasticity. As a consequence, the implementation generated by \sys will likely need to change over time. While \sys's architecture is designed to tackle that aspect by not having hard-wired rules for code generation, we will also devise new runtime monitoring and adaptive code generation techniques, in the spirit of prior work~\cite{pyxis,eddies,aqp-survey}.


\section{Conclusion}
\label{sec:conclusion}
We are optimistic that the time is ripe for innovation and adoption of new technologies for end-user distributed programming. This is based not only on our assessment of research progress and potential, but also the emerging competition among cloud vendors to bring third-party developers to their platforms. We are currently implementing the \runtime runtime, and exploring different algorithms to lift legacy design patterns to \logic. Our next goals are to design the compilation strategies from \logic to \runtime programs, and to explore the compilation of application-specific availability and consistency protocols.
We are also contemplating related research agendas in security and developer experience including debugging and monitoring.
There are many research challenges ahead, but we believe they can be addressed incrementally and in parallel, and quickly come together in practical forms.

\section{Acknowledgments}
This work is supported in part by the National Science Foundation through grants CNS-1730628, IIS-1546083, IIS-1955488, IIS-2027575, CCF-1723352, and DOE award DE-SC0016260; the Intel-NSF CAPA center, and gifts from Adobe, Amazon, Ant Group, Ericsson, Facebook, Futurewei, Google, Intel, Microsoft, NVIDIA, Scotiabank, Splunk and VMware.








\scriptsize
\bibliographystyle{abbrv}
\bibliography{confs,akcheung,extrarefs}
\normalsize
\appendix
\section{Lifting Legacy Design Patterns}
\label{sec:lifting-appendix}
We aim for \logic to be a general-purpose IR that can be targeted from a range of input languages.
In this section, we provide initial evidence that \logic can be a natural target for code written in a variety of accepted distributed design patterns: actors, futures, and MPI. We provide working implementations of all code from the paper at \url{https://github.com/hydro-project/cidr2021}.



\subsection{Actors}
The Actor model has a long history~\cite{hewitt1977viewing}. In a nutshell, an actor is an object with three basic primitives~\cite{agha1990concurrent}: (a) exchange messages with other actors, (b) update local state, and (c) spawn additional actors. Like other object abstractions, actors have private, encapsulated state. Actors are often implemented in a very lightweight fashion running in a single OS process; actor libraries like Erlang can run hundreds of thousands of actors per machine. At the same time, the asynchronous semantics of actors makes it simple to distribute them across machines.

Actors are like objects: they encapsulate state and handlers. \logic does not bind handlers to objects, but we can enforce that when lifting by generating a \logic program in which we have an \texttt{Actor} class keyed by \texttt{actor\_id}, and each handler's first argument identifies an \texttt{actor\_id} that associates the inbound message with a particular \texttt{Actor} instance. The \logic to emulate spawning an actor simply creates a new \texttt{Actor} instance with a unique ID and runs any initialization code to associate initial state with the actor. In keeping with other actor implementations, each actor is very lightweight. \logic allows us to optionally specify availability, consistency and deployment for our actors' handlers. \compiler can choose to how to partition/replicate actors across machines.

Actor frameworks provide event loops, and at first blush it is straightforward to map an actor method into \logic. Consider an actor method \texttt{do\_foo(msg)} with an RPC-like behavior in Erlang style:
\vspace{-0.5em}
\begin{plainlisting}
do_foo(msg) ->
  foo(msg);
\end{plainlisting}
\vspace{-1em}
This translates literally into a \logic handler:
\begin{hydrolisting}{A Simple Actor Method in \logic}
on do_foo(actor_id, msg): 
  return foo(msg)
\end{hydrolisting}
RPC is a good match to the transducer model where code fragments are bracketed by send and receive. 
But actors are not transducers. In particular, they can issue blocking requests for messages at any point in their control flow. In the next listing, note that the actor first runs the function \texttt{m\_pre(msg)}, then waits to receive a message in the \texttt{mybox} mailbox, after which it runs \texttt{m\_post} on the message it finds:
\vspace{-0.5em}
\begin{plainlisting}
m(msg) ->
  m_pre(msg)
  receive
    {mybox, newmsg} ->
      m_post(newmsg)
  end.
\end{plainlisting}
\vspace{-1em}
We can translate this into two separate handlers in \logic, but we need to make sure that (a) the state of the computation (heap and stack) after \texttt{m\_pre} runs is preserved, (b) \texttt{m\_post} can run from that same state of computation, and (c) that the handler doesn't do anyting else while waiting for \texttt{newmsg}. \emph{Coroutines} are a common language construct that provides convenient versions of (a) and (b), and are found in many host languages for actors (including C\# for Orleans and Scala for Akka). The third property (c) can be enforced across \logic ticks by a status variable in the actor\footnote{The attentive reader will note that we have elided a bit of bookkeeping here that buffers new inbound messages to \texttt{m} while the actor is waiting.}:
\begin{hydrolisting}{Mid-Method Message Handling in \logic}
on m(actor_id: int, msg):
  actors[actor_id].state := m_pre(msg)
  actors[actor_id].waiting := true
on m_receive_mybox(actor_id: int, newmsg): 
  result = m_post(actors[actor_id].state, newmsg)
  actors.delete([actor_id])
  return result
\end{hydrolisting}
Note that this \logic has to use non-monotonic mutation to capture the (arguably undesirable!) notion of blocking that is implicit in a synchronous receive call.

\subsection{Promises and Futures}
Another common language pattern for distributed messaging is Promises and Futures; this has roots in the actor literature as well~\cite{baker1977incremental}, but often appears independently. The basic idea is to spawn an asynchronous function call with a handle for the computation called a \emph{Promise}, and a handle for the result called a \emph{Future}.  In the basic form, sequential code generates pairs of Promises and Futures, effectively launching the computation of each Promise in a separate thread of execution (perhaps on a remote machine), and continuing to process locally until the Future needs to be resolved. We take an example from the Ray framework in Python:
\vspace{-0.4em}
\begin{plainlisting}
futures = [f.remote(i) for i in range(4)]
x = g()
print(ray.get(futures))
\end{plainlisting}
\vspace{-1em}
The function \texttt{f} is invoked as a promise for the numbers 0 through 3 via Ray's \texttt{f.remote} syntax; four futures are immediately stored in the array \texttt{futures}. The function \texttt{g()} then runs locally while the promises execute concurrently and remotely. After \texttt{g()} completes, the futures are resolved (in batch, in this case) by the \texttt{ray.get} syntax. In this simple example, futures are little more than a syntactic sugar for keeping track of asynchronous promise invocations. The translation to \logic is straightforward. It could be sugared similarly to Ray if desired, but we show it in full below.
Much like our mid-method receipt for actors, we implement waiting across \logic ticks with a condition variable.

\begin{hydrolisting}{Promises/Futures in \logic}
import promises from PromisesEngine
var waiting
on start: |\label{line:rayexamplestart}|
  send promises(handle: int, f, i: int):
    {(unique_id(), f, i) for i in range(4)})
  x := g() 
  waiting := true
on futures(handle: int, result).len() >= 4:
  print([f.result for f in futures])
  futures.delete()
  waiting := false |\label{line:rayexampleend}|
\end{hydrolisting}



Promise/Future libraries vary in their semantics, and it's relatively easy to generate \logic code for each of these semantics. For example, note that promises and futures are data, so we can implement semantics where they can be sent or copied to different logical agents (like our actors above).
Similarly, we can support a variety of ``kickoff'' semantics for promises. Our example above eagerly executes promises, but we could easily implement a lazy model, where pending promises are placed in a table until future requests come along.

\subsection{MPI Collective Communications}
MPI is a standard Message Passing Interface for scientific computing and supercomputers~\cite{mpi2}. While this domain per se is not a primary target for \logic, the ``collective communication'' patterns defined for MPI are a good suite of functionality that any distributed programming environment should support. 

The MPI standard classifies these patterns into the following categories:

    \smallitem{One-to-All}, in which one agent contributes to the result, but all agents receive the result. The two basic patterns are \texttt{Bcast} (which takes a value and sends a copy of it to a set of agents) and \texttt{Scatter} (which takes an array and partitions it, sending each chunk to a separate agent).
    
    \smallitem{All-to-One}, in which all agents contribute to the result, but one agent receives the result. The basic patterns are \texttt{Gather} (which assembles data from all the agents into a dense array) and \texttt{Reduce} (which computes an aggregate function across data from all agents).
    
    \smallitem{All-to-All}, in which all agents contributes to \emph{and} receive the result. This includes \texttt{Allgather} (similar to gather but all agents receive all the data and assemble the result array), \texttt{Allreduce} (similar to reduce except the result arrives at all agents) and \texttt{Alltoall} (all processes send and receive the same amount of data to each other).
\smallitembot

These operations map very naturally into \logic. Assume we start with a static table \texttt{agents} containing the set of relevant agentIDs.
\begin{hydrolisting}{MPI collective communication in \logic}
table agents(agent_id: int, key=agent_id)
query acount int:
  agents.count()
table gathered(request_id: int, ix: int, val, 
               tombstone: bool,key=(request_id, ix))
query gcount(req_id: int, cnt: int):
  (req_id, gathered[req_id].count())
  
on mpi_bcast(msg_id: int, msg):
  send mpi_bcast_channel(agent_id: int, msg_id, int, msg):
    {(a.aid, msg_id, msg} for a in agents))
    
on mpi_scatter(req_id: int, arr):
  send mpi_scatter_channel(agent_id: int, req_id: int, 
                           subarray):
    chunksz = arr.len()/acount
    if chunksz > 1:
      {(i, req_id, arr[range(i*chunksz, (i+1)*chunksz - 1)]) 
       for i in acount }
    else:
      {(i, req_id, arr[i]) for i in arr.len() }
    
on mpi_gather(req_id: int, ix: int, val):
  gathered.merge(req_id, ix, val)
  if (gcount[req_id] >= acount):
    result = gathered[req_id].array_agg(val, order=ix)
    gathered[req_id].tombstone.merge(true)
    return result
    
on mpi_reduce(req_id: int, ix: int, val, lambda):
  gathered.merge(req_id, ix, val)
  if (gcount[req_id] >= acount):
    result = reduce(lambda, gathered[req_id])
    gathered[req_id].tombstone.merge(true)
    return result

on mpi_allgather(req_id: int, ix: int, val):
  result = mpi_gather(req_id, ix, val)
  if (result):
    send mpi_bcast(req_id, result)
    
on mpi_allreduce(req_id: int, ix: int, val, lambda):
  result = mpi_reduce(req_id, ix, val)
  if (result):
    send mpi_bcast(req_id, result)
\end{hydrolisting}
Note that these are naive specifications, and there are various well-known optimizations that can be employed by \compiler, including tree-based or ring-based mechanisms.



\end{document}